\documentclass[pra,twocolumn,amsmath,amssymp,%showpacs, %showkeys, 
							nofootinbib]{revtex4}
\usepackage{graphicx}   % Include figure files
\usepackage{dcolumn}
\usepackage{bm}         % bold math
\newcommand{\rd}{{\rm d}}
\newcommand{\re}{{\rm e}}
\newcommand{\ri}{{\rm i}}

\newcommand{\mB}{\mu_{\rm B}}
\begin{document}

\title{Dressed matter waves} 

\author{Andr\'{e} Eckardt}

\author{Martin Holthaus}

\email{holthaus@theorie.physik.uni-oldenburg.de}

\affiliation{Institut f\"ur Physik, Carl von Ossietzky Universit\"at, 
		D-26111 Oldenburg, Germany}
              
\date{January 1, 2008}

\begin{abstract}
We suggest to view ultracold atoms in a time-periodically shifted optical
lattice as a ``dressed matter wave'', analogous to a dressed atom in an
electromagnetic field. A possible effect lending support to this concept 
is a transition of ultracold bosonic atoms from a superfluid to a 
Mott-insulating state in response to appropriate ``dressing'' achieved
through time-periodic lattice modulation. In order to observe this effect 
in a laboratory experiment, one has to identify conditions allowing for
effectively adiabatic motion of a many-body Floquet state.  
\end{abstract}

\maketitle
%\pacs{03.75.Lm, 42.50.Hz, 64.70.Tg}

%%%%%%%%%%%%%%%%%%%%%%%%%%%%%%%%%%%%%%%%%%%%%%%%%%%%%%%%%%%%%%%%%%%%%%%%%%%%%%%%

\section{Introduction}
\label{S_1}

The ``dressed-atom picture'' provides a transparent approach to the dynamics of 
atoms and molecules in strong electromagnetic fields~\cite{CohenTannoudji04}.
In essence, the ``dressing''  provided by the field may equip the atom or 
molecule with properties quite different from those of a ``bare'' one. 
A hallmark example along these lines is given by the modification of Zeeman 
hyperfine spectra of atoms interacting with strong radiofrequency fields: 
As reviewed in Sec.~\ref{S_3} below, in the presence of the dressing field 
the bare atomic $g$-factors become multiplied by a certain Bessel function, 
the argument of which is proportional to the strength of the field, and 
inversely proportional to its frequency. Thus, experimentally recorded 
hyperfine spectra depend sensitively on these parameters~\cite{HarocheEtAl70}. 
Closely related phenomena have been observed, for instance, in radiatively 
assisted collisions of Rydberg atoms~\cite{ThomsonEtAl92}.

The appearance of a Bessel function in response to time-periodic forcing 
is typical for quantum systems that can be viewed as a set of interacting 
nearest neighbours, such as the angular momentum substates in the case of 
the $g$-factor modification. Another striking example has emerged only 
recently: In experiments with ultracold atoms in time-periodically shifted 
optical lattices, the hopping matrix elements which quantify the magnitude 
of the tunnelling contact between states located at adjacent lattice sites 
differ from those of a bare lattice system again by a Bessel function, and 
thus can be tuned by adjusting the amplitude or the frequency of the lattice 
modulation~\cite{LignierEtAl07,SiasEtAl07}.  This finding now suggests an 
interesting question: The $g$-factor experiment~\cite{HarocheEtAl70} with 
single atoms has been instrumental for establishing the dressed-atom picture; 
could the recent experiments~\cite{LignierEtAl07,SiasEtAl07} with  
Bose--Einstein condensates lead to a similar picture of ``dressed matter 
waves''? In other words, can one exploit time-periodic forcing for endowing 
a macroscopic matter wave with properties it does not have when the forcing 
is absent? 

In order to expand on this question, we proceed as follows: We first recall
in Sec.~\ref{S_2} the physics underlying the Bessel-function modification
occurring in the dressed-atom picture, using the example of a two-level system 
interacting with a quantised radiation mode. However, when dealing with 
cold atoms in time-periodically shifted optical lattices, it is certainly 
reasonable to describe the time-periodic lattice modulation in terms
of an external classical force. Hence, we discuss in Sec.~\ref{S_3} how the
Bessel function appears in that context, employing the Floquet formalism.
For fully exploiting the possibilities of control opened up by time-periodic
forcing, adiabatic response to slowly changing parameters plays an important
role. Therefore, we briefly point out in Sec.~\ref{S_4} how the adiabatic 
principle works for Floquet states. In Sec.~\ref{S_5} we put all these 
pieces together and argue that a system of interacting ultracold bosonic atoms 
in a time-periodically modulated optical lattice can change its state from 
superfluid to Mott-insulator-like and back, if the modulation is switched on 
and off in an adiabatic manner, {\em and\/} the parameters are chosen 
judiciously~\cite{EckardtEtAl05,EckardtHolthaus07,CreffieldMonteiro06}. This 
scenario, still to be confirmed experimentally, could lend significant support 
to the notion of dressed matter waves: The dressing achieved through the 
lattice modulation determines the state of the system. We finally sum up our
conclusions in Sec.~\ref{S_6}.

\section{The dressed two-level system}
\label{S_2}

We start by studying a ``two-level atom'' interacting with both a static 
external field and a single mode of a quantised radiation field. Its dynamics 
are governed by the Hamiltonian
\begin{equation}
	H = H_{\rm at} + H_{\rm int} + H_{\rm rad} \; ,
\label{eq:HAM}
\end{equation} 
where
\begin{equation}
	H_{\rm at} = -\frac{J}{2}\left( \begin{array}{rr}
		0 & 1 \\
		1 & 0 \end{array} \right)
\label{eq:ATM}
\end{equation}
specifies the unperturbed system with energy eigenvalues $\pm J/2$,
\begin{equation}
	H_{\rm rad} = \hbar\omega\left( -\frac{1}{2}\partial_z^2 
	+ \frac{1}{2} z^2 \right) \left( \begin{array}{rr}
		1 & 0 \\
		0 & 1 \end{array} \right)
\end{equation}
models a radiation mode with frequency $\omega$ in terms of a harmonic
oscillator with dimensionless oscillator coordinate~$z$, and		
\begin{equation}
	H_{\rm int} = \frac{1}{2}\left( K_0 + \gamma z \right) 
	\left( \begin{array}{rr}
		1 &  0 \\
		0 & -1 \end{array} \right)
\end{equation}
describes the static field of strength $K_0$ and the coupling to the 
radiation mode, with a strength specified by a constant $\gamma$. When 
$J = 0$, so that the two ``atomic'' levels are degenerate, this 
Hamiltonian~(\ref{eq:HAM}) obviously is diagonalised by the shifted 
harmonic-oscillator states 
\begin{eqnarray}
	\psi_{n,+}(z) &=& \left( \begin{array}{c}
		\varphi_n(z + \gamma/2\hbar\omega) \\
		0	\end{array} \right) \;,
	\nonumber\\
	\psi_{n,-}(z) &=& \left( \begin{array}{c}
		0				  \\
		\varphi_n(z - \gamma/2\hbar\omega) 
			\end{array} \right) 	
\label{eq:UPS}
\end{eqnarray}
with energies
\begin{equation}
	E_{n,\pm}^{(0)} = \hbar\omega\!\left( n + \frac{1}{2} \right) 
	\pm \frac{K_0}{2} - \frac{\gamma^2}{8\hbar\omega} \; ;
\end{equation}	 
the functions $\varphi_n(z)$ denote the familiar eigenfunctions of the 
harmonic oscillator,
\begin{equation}
	\varphi_n(z) = (\sqrt{\pi} \, 2^n \, n!)^{-1/2} \, 
	H_n(z) \, \exp(-z^2/2) \; .
\end{equation}
The static field splits the two atomic levels by the amount $\Delta E = K_0$.
When this splitting is matched by an integer number of photons, that is, when
\begin{equation}
	K_0 = (\ell - n) \hbar\omega \; ,
\label{eq:RES}
\end{equation}
the unperturbed states are pairwise degenerate, 
$E_{n,+}^{(0)} - E_{\ell,-}^{(0)} = 0$. 
We now study the removal of this degeneracy between an ``$n$-photon state''
and an ``$\ell$-photon state'' for nonzero $J$, assuming $J \ll \hbar\omega$.

Degenerate-state perturbation theory requires to evaluate the matrix
elements of the ``perturbation'' $H_{\rm at}$ in the basis~(\ref{eq:UPS}), 
and, hence, to compute the overlap integrals
\begin{equation}   	
	M = \int_{-\infty}^{+\infty} \! \rd z \, 
	\varphi_n(z + \gamma/2\hbar\omega) \, 
	\varphi_\ell(z - \gamma/2\hbar\omega) \; .
\end{equation}
With the help of the expansion
\begin{equation}
	H_n(z + \beta) = \sum_{k=0}^n 
		{n \choose k} H_k(z) \, (2\beta)^{n-k}
\end{equation}
this integral can be calculated exactly, yielding
\begin{equation}
	M = \exp(-\alpha/2) \, \alpha^{(n-\ell)/2} \, \sqrt{n! \, \ell!}
	\sum_{k=0}^{\min(n,\ell)} \!\!
	\frac{(-\alpha)^{\ell-k}}{k! \, (n-k)! \, (\ell-k)!} \; ,
\label{eq:EXP}
\end{equation} 		
where we have introduced the dimensionless parameter
\begin{equation}
	\alpha = 2 \left( \frac{\gamma}{2\hbar\omega} \right)^2 \; .
\label{eq:ALP}
\end{equation}		
Assuming $\ell \ge n$, and employing the generalised Laguerre 
polynomials~\cite{AbramowitzStegun72}
\begin{equation}
	L_n^{(\ell - n)}(x) = \sum_{k = 0}^n
	\frac{\ell! \, (-x)^k}{k! \, (n - k)! \, (\ell - n + k)!} \; ,
\end{equation}
this expression~(\ref{eq:EXP}) takes the form
\begin{equation}
	M = (-1)^{\ell - n} \exp(-\alpha/2) \, \alpha^{(\ell-n)/2} \,
	\sqrt{\frac{n!}{\ell!}} \, L_n^{(\ell - n)}(\alpha) \; . 
\end{equation}
Now we are interested in the limiting case of almost classical fields
containing a very large number of photons. In order to maintain the
resonance condition~(\ref{eq:RES}), we keep the integer $\nu \equiv \ell - n$
fixed while letting $n$ and $\ell$ tend to infinity. In that limit, one 
has~\cite{AbramowitzStegun72}
\begin{eqnarray}
	L_n^{(\ell - n)}(x) 
	&\stackrel{n \to \infty, \, \ell - n = \nu}{\longrightarrow}&
	\frac{\ell!}{n!} \, \re^{x/2} \,  
	\left(\frac{n+\ell+1}{2} x\right)^{-(\ell-n)/2} \,
	\nonumber\\&&\times\;
	{\rm J}_{\ell - n}\!\left( \sqrt{2(n+\ell+1)x} \right)  ,	  
\end{eqnarray}
where ${\rm J}_\nu(x)$ denotes a Bessel function of integer order~$\nu$.
Hence, one finally obtains
\begin{equation}
	M \, \longrightarrow \, (-1)^{\ell - n}
	{\rm J}_{\ell - n}\!\left( \sqrt{2(n+\ell+1)\alpha} \right) 
\end{equation}
in that same limit. 

It remains to interpret the peculiar-looking argument of the Bessel function. 
When placing the field oscillator into a coherent state, its amplitude $z_0$ 
is determined by energy considerations: Since, as expressed by the resonance 
condition~(\ref{eq:RES}), the field is exchanging $\ell - n$ photons with 
the atom, the average field energy is the arithmetic mean of the energy of 
an $n$-photon state and that of an $\ell$-photon state. This gives       
\begin{equation}
	\frac{1}{2}\left(n + \frac{1}{2} \; + \; \ell + \frac{1}{2}\right)
	= \frac{1}{2} z_0^2 \; ,
\end{equation}
which, in view of the definition~(\ref{eq:ALP}), implies
\begin{equation}
	\sqrt{2(n+\ell+1)\alpha} = \frac{\gamma z_0}{\hbar\omega} \; .
\end{equation}

The energy eigenvalues which have been degenerate for $J = 0$, {\em i.e.\/},
$E_{n,+}^{(0)} = E_{\ell,-}^{(0)}$, now are shifted by $\pm MJ/2$ for 
nonzero~$J$. For energies sufficiently high to validate the preceding
reasoning, the spectrum of the Hamiltonian~(\ref{eq:HAM}) thus consists
of a sequence of doublets split by $|MJ|$, the doublet centers being 
separated by $\hbar\omega$. Putting all things together, this means 
that the ``atom'' $H_{\rm at}$, when ``dressed'' by the interaction 
$H_{\rm int} + H_{\rm rad}$ under the conditions detailed above, behaves 
like a {\em noninteracting\/} system~(\ref{eq:ATM}) with a modified level 
splitting determined by the effective $J$-parameter 
\begin{equation}
	J_{\rm eff} = (-1)^\nu 
	{\rm J}_\nu\!\left(\frac{\gamma z_0}{\hbar\omega}\right) J \; .
\label{eq:JZS}
\end{equation}
This is the lesson to be learned from the present two-level example: When the 
driving field can be considered classical, resonant forcing effectuates a
modification of the ``atomic'' level splitting such that the unperturbed 
splitting is multiplied by a Bessel function with an argument proportional 
to the driving amplitude divided by $\hbar\omega$. The order $\nu$ of this 
Bessel function is determined by the number of photons in resonance with the 
transition, according to Eq.~(\ref{eq:RES}). In particular, when there is no 
static field, one has $\nu = 0$ and thus recovers the modification of the
splitting by a $J_0$ Bessel function which also underlies, for instance, the 
coherent destruction of tunnelling of a single particle in driven symmetric
double well potential~\cite{GrossmannEtAl91,LlorentePlata92}. Quite recently,
this phenomenon has been observed with cold atoms in periodic double-well
potentials~\cite{Oberthaler07}.

\section{Elements of Floquet theory}
\label{S_3}

In order to avoid the consideration of a quantised field and to start 
with a classical driving force right away, we now treat the explicitly 
time-dependent Hamiltonian  
\begin{equation}
	H(t) = H_0 + H_1(t) \; ,
\label{eq:HTL}	
\end{equation}	
where the time-independent system $H_0$ corresponds to a spin~1 in a magnetic
field~$B$ oriented in the $x$-direction,
\begin{equation}
	H_0 = g_1\mB B \, \frac{1}{\sqrt{2}}\left(
	\begin{array}{ccc}
        	0          & 1            & 0            \\
		1          & 0            & 1            \\
		0          & 1            & 0            \\
	\end{array} \right) \; ,
\label{eq:TLS}
\end{equation}
with $\mB$ denoting the Bohr magneton, and $g_1$ the Land\'e $g$-factor.
The external forcing is given as an additional static magnetic field $B_0$ 
and an oscillating field with amplitude $B_\omega$ and frequency~$\omega$, 
both directed along the $z$-axis,	
\begin{equation}	
	H_1(t) = g_1\mB  \big[B_0 + B_\omega\cos(\omega t) \big]
	\left(\begin{array}{rrr}
        	1       & 0            & 0            \\
        	0       & \phantom{-}0 & 0            \\
		0       & 0            & -1           \\
	\end{array} \right) \; .
\label{eq:TLF}
\end{equation}	
Apart from the fact that here the forcing is truly classical, this system 
closely resembles the previous two-level example~(\ref{eq:HAM}): The
unperturbed system is characterised by ``nearest-neighbour coupling'', 
while the forcing is diagonal. Now the Hamiltonian~(\ref{eq:HTL}) depends
periodically on time, 		
\begin{equation}
	H(t) = H(t + T) \; ,
\end{equation}	
with period $T = 2\pi/\omega$. Hence, the Schr\"odinger equation
\begin{equation}
	\ri\hbar \frac{\partial}{\partial t} |\psi(t) \rangle
	= H(t) \, |\psi(t) \rangle
\end{equation}
has Floquet-type solutions~\cite{AutlerTownes55,Shirley65,Zeldovich67,Ritus67}
\begin{equation}
	|\psi_\alpha(t)\rangle = |u_\alpha(t)\rangle
	\exp(-\ri\varepsilon_\alpha t/\hbar) \; ,
\end{equation}
where the functions $|u_\alpha(t)\rangle$ inherit the periodic nature
of $H(t)$,
\begin{equation}	
	|u_\alpha(t)\rangle = |u_\alpha(t+T)\rangle \; .
\end{equation}
These functions, together with the corresponding quasienergies 
$\varepsilon_\alpha$, are obtained as solutions to the eigenvalue problem
\begin{equation}
	\big( H(t) - \ri\hbar\partial_t \big) |u_\alpha(t)\rangle\!\rangle 
	= \varepsilon_\alpha \, |u_\alpha(t)\rangle\!\rangle \; , 
\label{eq:QEE}
\end{equation}
which is defined in an {\em extended Hilbert space\/} 
$\mathcal{H} \otimes \mathcal{T}$ of $T$-periodic functions~\cite{Sambe73}
in which the time~$t$ is regarded as a {\em coordinate\/} and which, 
therefore, is equipped with the scalar product		
\begin{equation}
	\langle\!\langle \, \cdot \, | \, \cdot \, \rangle\!\rangle
	\equiv \frac{1}{T} \int_0^T \! \rd t \,
	\langle \, \cdot \, | \, \cdot \, \rangle \; , 
\label{eq:SKP}
\end{equation}
combining the standard scalar product 
$\langle \, \cdot \, | \, \cdot \, \rangle$ for the system's original Hilbert 
space $\mathcal{H}$ with time-averaging. We stick to the convention of writing 
$|u_\alpha(t)\rangle$ for a Floquet function viewed in $\mathcal{H}$, but 
$|u_\alpha(t)\rangle\!\rangle$ when that same function is regarded as an 
element of the extended space $\mathcal{H} \otimes \mathcal{T}$.
  
There is one issue implied by the Floquet formalism which requires particular 
attention. Namely, if $| u_{(n,0)}(t)\rangle\!\rangle$ solves
\begin{equation}
	\big( H(t) - \ri\hbar\partial_t \big) |u_{(n,0)}(t)\rangle\!\rangle
	= \varepsilon_{(n,0)} |u_{(n,0)}(t)\rangle\!\rangle 	
\end{equation}
with quasienergy $\varepsilon_{(n,0)}$, then
\begin{equation}
	| u_{(n,m)}(t)\rangle\!\rangle
	\equiv | u_{(n,0)}(t)\rangle\!\rangle \exp(\ri m \omega t)
\end{equation}
solves
\begin{equation}	
	\big( H(t) - \ri\hbar\partial_t \big) |u_{(n,m)}(t)\rangle\!\rangle
	= \varepsilon_{(n,m)} |u_{(n,m)}(t)\rangle\!\rangle	
\end{equation}
with quasienergy
\begin{equation}
	\varepsilon_{(n,m)} = \varepsilon_{(n,0)} + m \hbar\omega \; ,
\end{equation}
where $m$ is {\em any} (positive or negative) integer. Hence, the 
quasienergy spectrum repeats itself periodically on the energy-axis;
each ``Brillouin zone'' of width $\hbar\omega$ contains one respresentative,
labelled by~$m$, of the class of eigenvalues belonging to the Floquet state
labelled by~$n$. But when following the evolution of a wave function 
$|\psi(t)\rangle$ in the physical Hilbert space $\mathcal{H}$, only {\em one\/}
representative from each class is needed, giving an expansion of the form
\begin{equation}
	|\psi(t)\rangle = \sum_n c_n \, | u_{(n,0)}(t)\rangle \,  
	\exp(-\ri\varepsilon_{(n,0)}t/\hbar)
\end{equation}
with time-independent coefficients~$c_n$.

In order to apply this lore to the spin-1-system~(\ref{eq:HTL}), we observe
that the Floquet basis states
\begin{eqnarray}
	|u_{(+,m)}(t)\rangle\!\rangle & = &
    	\left(\begin{array}{c} 1 \\ 0 \\ 0 \\ \end{array} \right)
    	\exp\!\left(-\ri\frac{g_1\mB B_\omega}{\hbar\omega}\sin(\omega t) 
    	+ \ri m \omega t\right)
\nonumber  \\
	|u_{(0,m)}(t)\rangle\!\rangle & = &
    	\left(\begin{array}{c} 0 \\ 1 \\ 0 \\ \end{array} \right)
    	\exp(\ri m \omega t)
\nonumber \\
	|u_{(-,m)}(t)\rangle\!\rangle & = &
    	\left(\begin{array}{c} 0 \\ 0 \\ 1 \\ \end{array} \right)
    	\exp\!\left(+\ri\frac{g_1\mB B_\omega}{\hbar\omega}\sin(\omega t) 
    	+ \ri m \omega t\right)
\nonumber\\
\end{eqnarray}
diagonalise the quasienergy operator $H_1(t) - \ri\hbar\partial_t$ which
is obtained when there is no field~$B$, so that the eigenstates of the 
three-level Hamiltonian~(\ref{eq:TLS}) are degenerate; the ``unperturbed'' 
quasienergies express the Zeeman splitting caused by the other static
field~$B_0$:  	 
\begin{eqnarray}
	\varepsilon^{(0)}_{(n,m)} = n \cdot g_1\mB B_0 + m \hbar\omega
	\qquad (n = 0,\pm 1) \; .
\end{eqnarray}
If the oscillating field is resonant in the sense that 
\begin{equation}
	g_1 \mB B_0 = \nu \hbar \omega \; ,
\label{eq:REB}
\end{equation}
then the Floquet functions 
\begin{eqnarray}
	|u_1(t)\rangle\!\rangle & \equiv & |u_{(+,0)}(t)\rangle\!\rangle
\nonumber \\ 
	|u_2(t)\rangle\!\rangle & \equiv & |u_{(0,\nu)}(t)\rangle\!\rangle
\nonumber \\
	|u_3(t)\rangle\!\rangle & \equiv & |u_{(-,2\nu)}(t)\rangle\!\rangle
\end{eqnarray}
correspond to the same quasienergy, and thus are degenerate. The removal
of this degeneracy for nonvanishing~$B$ once again is assessed by
degenerate-state perturbation theory, assuming $g_1 \mB B \ll \hbar\omega$. 
In contrast to Sec.~\ref{S_2}, now the calculation proceeds in the extended 
Hilbert space $\mathcal{H} \otimes \mathcal{T}$, and thus invokes the 
computation of matrix elements 
$\langle\!\langle u_j | H_0 | u_k \rangle\!\rangle$ with respect to the 
scalar product~(\ref{eq:SKP}). But this is what makes the mathematics quite 
simple: Using the identity
\begin{equation}
	\re^{\ri z \sin\omega t} = \sum_{k=-\infty}^{+\infty} 
	\re^{\ri k \omega t} {\rm J}_k(z)  
\end{equation}	
for expanding the unperturbed Floquet functions, time averaging according to 
the definition~(\ref{eq:SKP}) serves to filter out one particular term from 
the sum, determined by the resonance condition~(\ref{eq:REB}). Thus, one 
immediately obtains
\begin{equation}
	\langle\!\langle u_j | H_0 | u_k \rangle\!\rangle
	= (-1)^\nu g_1 \mB B \, {\rm J}_\nu \!
	\left(\frac{g_1\mB B_\omega}{\hbar\omega} \right)
	\cdot \frac{1}{\sqrt{2}} \delta_{j,k\pm 1} \; ,
\end{equation}
giving the quasienergies
\begin{equation}
	\varepsilon_{(n,m)} = n \cdot (-1)^\nu g_1 \mB B \, {\rm J}_\nu \!	
	\left(\frac{g_1\mB B_\omega}{\hbar\omega} \right) + m \hbar\omega \; . 
\end{equation}
Hence, the effect of the forcing~(\ref{eq:TLF}) on the system~(\ref{eq:TLS}) 
is described by replacing the bare $g$-factor $g_1$ by the effective
substitute  
\begin{equation}
	g_{\rm eff} = (-1)^\nu {\rm J}_\nu \!
	\left(\frac{g_1\mB B_\omega}{\hbar\omega} \right) g_1 \; .
\label{eq:JZB}
\end{equation}  
Evidently, the line of reasoning adopted in this section to treat classical
forcing parallels the arguments given in Sec.~\ref{S_2} for a system
interacting with a quantised field. But here the argument is considerably more
direct, avoiding the analysis referring to ``large photon numbers''. The price
to pay for this simplification is a quasienergy spectrum which is strictly
$\hbar\omega$-periodic und thus unbounded from below, whereas the exact 
quantum mechanical energy spectrum becomes approximately $\hbar\omega$-periodic
only for sufficiently high quantum numbers. Nonetheless, for systems subjected
to time-periodic classical forcing the Floquet picture combines great 
conceptual clarity with a  fairly succinct computational approach.

\section{Adiabatic following of Floquet states}
\label{S_4}

One additional piece of input is required before we can treat ultracold atoms
in a periodically shifted optical lattice, namely, the adiabatic response of 
Floquet states to slowly changing parameters. In order to make the point, 
let us briefly recapitulate the standard adiabatic theorem of quantum 
mechanics~\cite{BornFock28}: The task is solve a time-dependent Schr\"odinger 
equation 
\begin{equation}
	\ri \hbar \partial_t | \psi(t) \rangle = H^{P(t)} | \psi(t) \rangle
\label{eq:SGL}
\end{equation} 
with a Hamiltonian $H^{P(t)}$ depending on a parameter $P(t)$ which 
changes slowly in time. The strategy then is to ``freeze'' that parameter 
in a first step, and to consider the family of eigenvalue problems
\begin{equation}
	H^{P} | \varphi_n^P \rangle = E_n^P | \varphi_n^P \rangle
\end{equation}
for each relevant, fixed value of~$P$. Let us stipulate that the phases
of the instantaneous eigenstates $|\varphi_n^P \rangle$ be chosen such
that
\begin{equation}	 
	\langle \varphi_n^P | \partial_P \, \varphi_n^P \rangle = 0 \; .
\label{eq:PFC}
\end{equation}
If then the system initially, at time $t = 0$, is prepared in a particular
eigenstate, 
\begin{equation}
	|\psi(t=0)\rangle = |\varphi_n^{P(t=0)}\rangle \; ,
\end{equation}	 
and $P$ is allowed to vary sufficiently slowly, an approximate solution
to the Schr\"odinger equation~(\ref{eq:SGL}) is given by
\begin{equation}
	|\psi(t)\rangle = |\varphi_n^{P(t)}\rangle
	\exp\!\left(-\frac{\ri}{\hbar}\int_0^t \! \rd t' \, 
	E_n^{P(t')}\right) \; ,
\end{equation}	
provided the parameter variation proceeds smoothly, and $|\varphi_n^P \rangle$
is separated for all~$P$ by an energy gap from the other states. Hence, the
system stays in the state continuously connected to the one it was originally
prepared in, and acquires a ``dynamical'' phase determined by an integral
over the instantaneous energy eigenvalues encountered during its evolution.
We remark that it might not be possible to satisfy the phase-fixing 
condition~(\ref{eq:PFC}) globally if there is more than one time-dependent
parameter; this fact then forces one to explicitly introduce Berry's 
geometrical phase~\cite{Berry84}.  

When trying to transfer this adiabatic theorem to systems with a Hamiltonian
$H^{P(t)}(t)$ which would depend periodically on time if the parameter~$P$ 
where fixed, $H^{P}(t) = H^{P}(t+T)$, but which actually exhibits an 
additional ``slow'' time-dependence of $P$, one faces a problem: If one 
simply ``stopped the time'' in order to define an instantaneous Hamiltonian,
one would not only freeze the parameter~$P$, but also loose the periodic 
time-dependence. However, it appears much more natural to freeze {\em only\/} 
$P$, and to maintain the periodic time-dependence on the level of the 
instantaneous eigenvalue problems. The way to do so, as formulated in
Refs.~\cite{BreuerHolthaus89a,BreuerHolthaus89b}, includes a detour to the
extended Hilbert space $\mathcal{H} \otimes \mathcal{T}$ introduced in the 
previous section: Instead of starting from the actual Schr\"odinger equation
\begin{equation}
	\ri \hbar \partial_t | \psi(t) \rangle = 
	H^{P(t)}(t) | \psi(t) \rangle \; , 
\label{eq:HPT}
\end{equation}
one first distinguishes two different time variables, a variable $\tau$ 
for the slow, parametric time dependence and a variable $t$ for the
fast, oscillating one, and then considers the evolution equation   
\begin{equation}
	\ri \hbar \partial_{\tau} | \Psi(\tau,t) \rangle\!\rangle =
	\big( H^{P({\tau})}(t) - \ri \hbar \partial_t \big)
	| \Psi(\tau,t) \rangle\!\rangle   
\label{eq:EEE}
\end{equation}
in $\mathcal{H} \otimes \mathcal{T}$. If this equation can be solved,
one returns to the desired wave function $|\psi(t)\rangle$ evolving in
the system's true Hilbert space $\mathcal{H}$ by equating $\tau$ and $t$:
One has
\begin{equation}
	|\psi(t)\rangle = | \Psi(\tau,t) \rangle\!\rangle \Big|_{\tau = t}
	\; ,
\label{eq:PRO}
\end{equation}
since
\begin{eqnarray}
	\ri \hbar \partial_t | \psi(t) \rangle & = & 
	\ri \hbar \partial_\tau | \Psi(\tau,t) \rangle\!\rangle \Big|_{\tau = t}
      +	\ri \hbar \partial_t | \Psi(\tau,t) \rangle\!\rangle \Big|_{\tau = t}
\nonumber \\ 	& = & 
	\big( H^{P({\tau})}(t) - \ri \hbar \partial_t \big)
	| \Psi(\tau,t) \rangle\!\rangle \Big|_{\tau = t} 
\nonumber\\&&
      +\;	\ri \hbar \partial_t | \Psi(\tau,t) \rangle\!\rangle \Big|_{\tau = t}
\nonumber \\	& = &
	H^{P(t)}(t) | \psi(t) \rangle \; .
\end{eqnarray} 
On the level of the extended evolution equation~(\ref{eq:EEE}), one can 
now freeze~$P$ by stopping solely the time $\tau$, while leaving the other 
time~$t$ unaffected. This then defines the instantaneous eigenvalue problems 
in terms of the operators appearing on the right-hand side of 
Eq.~(\ref{eq:EEE}), 
\begin{equation}
	\big( H^{P}(t) - \ri \hbar \partial_t \big) 
	| u_\alpha^P(t)\rangle\!\rangle
	= \varepsilon_\alpha^P | u_\alpha^P(t) \rangle\!\rangle \; .
\end{equation}
Since, by construction, this problem lives in $\mathcal{H} \otimes \mathcal{T}$,
it is exactly the quasienergy problem formulated in Eq.~(\ref{eq:QEE}). The 
remaining reasoning follows the standard route: We fix the phases of the 
instantaneous eigenstates by requiring     
\begin{equation}	 
	\langle\!\langle u_\alpha^P | \partial_P \, u_\alpha^P 
	\rangle\!\rangle = 0 \; ,
\end{equation}
and start at time $\tau = 0$ with the initial condition 
\begin{equation}
	| \Psi(\tau\!=\!0,t) \rangle\!\rangle = 
	| u_\alpha^{P(\tau = 0)}(t) \rangle\!\rangle \; . 
\end{equation}
Then 
\begin{equation}
	| \Psi(\tau,t) \rangle\!\rangle =
	| u_\alpha^{P(\tau)}(t) \rangle\!\rangle
	\exp\!\left(-\frac{\ri}{\hbar}\int_0^\tau \! \rd \tau' \, 
	\varepsilon_\alpha^{P(\tau')}\right)
\end{equation}
is an adiabatic solution to the extended evolution equation~(\ref{eq:EEE}),
provided the propositions of the adiabatic theorem can be met, and returning 
to $\mathcal{H}$ according to Eq.~(\ref{eq:PRO}) gives
\begin{equation}
	| \psi(t) \rangle = | u_\alpha^{P(t)}(t) \rangle
	\exp\!\left(-\frac{\ri}{\hbar}\int_0^t \! \rd t' \,
	 \varepsilon_\alpha^{P(t')}\right)
\end{equation}
as an approximate solution to the original Schr\"odinger 
equation~(\ref{eq:HPT}). In short, for adiabatic quantum transport in 
periodically time-dependent systems with slowly changing parameters the 
Floquet states adopt a role which is completely analogous to that played 
by energy eigenstates in conventional situations described by an equation 
of the type~(\ref{eq:SGL}). The strategy of ``lifting'' the Schr\"odinger 
equation~(\ref{eq:SGL}) to the extended space $\mathcal{H} \otimes \mathcal{T}$,
applying standard techniques there, and then projecting back to $\mathcal{H}$ 
is useful not only for understanding the structure of the problem, but also for 
detailed computations of non-adiabatic corrections~\cite{DreseHolthaus99}.  

There is, however, a big caveat. As remarked above, the standard adiabatic 
theorem demands that the adiabatically transported state be separated by an
energy gap from all other states. Accordingly, when transferring this theorem 
to $\mathcal{H} \otimes \mathcal{T}$, one requires that the adiabatically 
transported Floquet state be separated in {\em quasienergy\/} from the other 
ones. But since one quasienergy-representative from each state falls into each
quasienergy Brillouin zone, this condition is almost impossible to satisfy 
when there is a large number of states. One then expects a multitude of 
near-degeneracies ``modulo $\hbar\omega$'', reflecting a dense set of
multiphoton resonances. In such a situation, it appears unlikely that an 
adiabatic limit exists~\cite{HoneEtAl97}. However, it appears equally plausibe       
that, if one does {\em not\/} consider the fictitious limit of a parameter
variation proceeding ``infinitely slowly'', but instead specifies that the
variation takes place within a finite time interval, most of these resonances 
are not ``seen'' long enough by the system to become active. Then effectively
adiabatic motion is possible, if major resonances can be avoided. Although it
might be hard to formulate this somewhat vague notion in a mathematically
precise manner in the general case, the emerging adiabatic principle (not 
theorem) for Floquet states can provide intuitively clear guidelines for
understanding the evolution of periodically driven systems in well-designed
particular cases. The following discussion of the driven Bose--Hubbard model 
exemplifies that the occurrence of effectively adiabatic motion, or its 
destruction by active resonances, depends on the choice of the frequency.

\section{The driven Bose--Hubbard model}
\label{S_5}

The Bose--Hubbard model, as devised by Fisher 
{\em et al.\/}~\cite{FisherEtAl89}, describes Bose particles on a lattice.
There exists a tunnelling contact between neigbouring sites, with a strength 
specified by a hopping matrix element~$J$; each pair of particles occupying 
the same site increases the energy of the system by an amount~$U$ due to
repulsion. Thus, for the case of a one-dimensional (1d) lattice with $M$ 
sites the many-body Hamiltonian reads   
\begin{equation}
	\hat{H}_0 = -J \sum_{\ell = 1}^{M-1}\left( 
	\hat{b}^\dagger_\ell \hat{b}^{\phantom\dagger}_{\ell+1} +
	\hat{b}^\dagger_{\ell+1} \hat{b}^{\phantom\dagger}_\ell \right)
	+ \frac{U}{2}\sum_{\ell=1}^M 
	\hat{n}_\ell \left( \hat{n}_\ell - 1 \right) \; ,
\label{eq:BHH}
\end{equation}	
where $\hat{b}^\dagger_\ell$ ($\hat{b}^{\phantom\dagger}_\ell$) is the creation
(annihilation) operator for a Bose particle at the $\ell$th lattice site, 
obeying $[\hat{b}^{\phantom\dagger}_\ell, \hat{b}^\dagger_k] = \delta_{\ell,k}$,
and $\hat{n}_\ell \equiv \hat{b}^\dagger_\ell \hat{b}^{\phantom\dagger}_\ell$ 
gives the number of particles on that site. Assuming that there are $N$ 
particles in total, and that the filling factor $n = N/M$ is integer, the 
system's ground state undergoes a significant change when the dimensionless 
control parameter $U/J$ is varied:  In the interaction-free limit $U/J \to 0$ 
it corresponds to a superfluid, given by a Bose--Einstein condensate with all 
particles occupying the lowest Bloch state,
\begin{equation}
	|{\rm SF} \rangle = \frac{1}{\sqrt{N!}}\left(\frac{1}{\sqrt{M}}
	\sum_{\ell=1}^M \hat{b}^\dagger_\ell \right)^N |0\rangle \; ,
\end{equation}	
where $|0\rangle$ is the vacuum state. In the opposite limit of vanishing 
tunnelling contact, $U/J \to \infty$, the individual sites are isolated, 
so that the systems adopts the Mott-insulating ground state
\begin{equation} 	
	|{\rm MI} \rangle = \prod_{\ell=1}^M
	\frac{(\hat{b}^\dagger_\ell)^n}{\sqrt{n!}}|0\rangle \; .	
\end{equation}
When the lattice is infinitely large, that is, for $M \to \infty$ and 
$N \to \infty$ while keeping $n = N/M$ constant at an integer value, a sharp 
transition between the superfluid and the Mott-insulating regime occurs at a 
critical value $(U/J)_{\rm c}$, accompanied by the emergence of a finite 
energy gap. For a 1d lattice with filling factor $n = 1$, one finds 
$(U/J)_{\rm c} \approx 3.4$~\cite{KuehnerEtAl00}. The Bose--Hubbard model 
has received considerable attention recently, since it can be realised 
with ultracold atoms in $d$-dimensional optical lattices 
($d = 1,2,3$)~\cite{JakschEtAl98}, allowing one to investigate the 
superfluid-to-Mott insulator quantum phase transition in great detail in 
the laboratory~\cite{GreinerEtAl02,StoeferleEtAl04,MunEtAl07,BlochEtAl07}.   

We view ultracold Bose particles in an optical lattice as prime candidates
for exploring the concept of dressed matter waves. Namely, atoms in a 1d 
lattice can be subjected to a time-periodic lattice modulation, to the effect 
that a term of the form
\begin{equation}
	\hat{H}_1(t) = \big[ K_0 + K_\omega \cos(\omega t) \big] 
	\sum_{\ell = 1}^M \ell \hat{n}_{\ell}
\label{eq:HDR}
\end{equation}   
is added to the system~(\ref{eq:BHH}). Here $K_\omega$ denotes the amplitude 
of a drive with angular frequency~$\omega$, typically on the order of one
to a few kilohertz~\cite{LignierEtAl07}, while $K_0$ corresponds to a static 
lattice tilt~\cite{SiasEtAl07}; the extension to lattices with higher 
dimension is straightforward. If there were no interaction between the 
particles, that is, for $U/J = 0$, the total Hamiltonian $
\hat{H}(t) = \hat{H}_0 + \hat{H}_1(t)$ would be identical in form to the 
systems studied in Secs.~\ref{S_2} and \ref{S_3}: One faces nearest-neighbour 
coupling combined with homogeneous site-diagonal forcing. Hence, when the 
resonance condition 
\begin{equation}
	K_0 = \nu \hbar\omega 
\end{equation}
corresponding to the previous equations~(\ref{eq:RES}) and (\ref{eq:REB})
is satisfied, so that the energy of $\nu$ ``photons'' matches the energy 
shift induced by the static tilt between adjacent sites, one can adapt the 
results~(\ref{eq:JZS}) and (\ref{eq:JZB}): Under such conditions, the driven 
system behaves approximately like an undriven one with the modified hopping 
matrix element
\begin{equation}
	J_{\rm eff} = (-1)^\nu {\rm J}_\nu(K_\omega/\hbar\omega) \, J \; .
\label{eq:MOD}
\end{equation}    
More careful analysis~\cite{EckardtEtAl05,EckardtHolthaus07} shows that this 
expression remains valid even for nonzero~$U$ at least in the high-frequency 
regime where $\hbar\omega \gg U$ and $\hbar\omega \gg J$, thus including the
strong-coupling case $U/J \gg 1$. The experimental data available so far give 
clear evidence of this modification~(\ref{eq:MOD}) both for $\nu = 0$, when 
there is no static tilt~\cite{LignierEtAl07}, and for $\nu = 1,2$, when one 
has ``photon''-assisted tunnelling~\cite{SiasEtAl07}. This finding now 
directly leads to a further consequence: The ratio $U/J$ governs the 
superfluid-to-Mott insulator transition in the bare Bose--Hubbard model; 
this control parameter has to be replaced by $U/J_{\rm eff}$ in the presence 
of resonant forcing. Since $J_{\rm eff}$ depends significantly on the 
amplitude~$K_\omega$, it should be possible to cross the border 
between the superfluid and the insulator regime by varying that 
amplitude~\cite{EckardtEtAl05,EckardtHolthaus07}. However, the notion of 
a ``superfluid'' or a ``Mott insulator'' refers to the ground state of the 
Bose--Hubbard model, so that it becomes necessary to guide the ground state 
of the undriven system $\hat{H}_0$ into the effective ground state of the 
driven system $\hat{H}_0 +  \hat{H}_1(t)$. That ``effective ground state'', 
of course, is the Floquet state which originates from the ground state of 
$\hat{H}_0$ when the drive is turned on, so that the adiabatic principle 
discussed in Sec.~\ref{S_4} comes into play: In an experiment aiming at 
a demonstration of a superfluid-to-Mott insulator transition induced by
time-periodic forcing, the driving amplitude should be turned on smoothly,
such that the system's wave function can follow the Floquet state connected
to the unperturbed ground state. But since adiabatic following in a 
periodically forced many-level system with a ``dense'' quasienergy spectrum 
is endangered by a host of multiphoton-like resonances, the precise choice 
of the protocol is not trivial: The frequency has to be chosen such that 
major resonances are avoided, while the amplitude has to vary sufficiently 
slowly in order to allow for a reasonable degree of adiabaticity, but still 
sufficiently fast in order to pass minor resonances before they become 
active. 

\begin{figure}[t]
\begin{center}
	\includegraphics[angle=-90.0, width =1\linewidth]{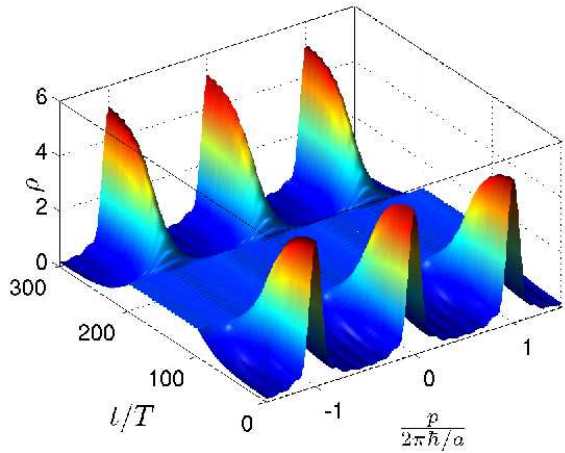}
\end{center}
\caption[FIG.~1]{Quasimomentum distribution computed according to
	Eq.~(\ref{eq:QMD}) for a driven Bose--Hubbard model with $N = 7$ 
	particles on $M = 7$ sites. The interaction strength is $U/J = 3$,
	the driving frequency is $\hbar\omega/J = 16$. Between $t = 0$ and 
	$t_1 = 100 \, T$ the amplitude $K_\omega(t)$ is increased 
	linearly from zero to $K_{\rm max} = 2.4\, \hbar\omega$, then kept 
	constant until $t_2 = 200 \, T$, and finally ramped linearly back 
	to zero between $t_2$ and $t_3 = 300 \, T$. Here the system is
	able to follow adiabatically.}
\label{F_1}
\end{figure}

We illustrate these deliberations by numerical calculations for small systems 
with $N = 7$ particles on $M = 7$ lattice sites. The initial $N$-body wave 
function $|\psi(0)\rangle$ at time $t = 0$ is chosen as the ground state of 
the Bose--Hubbard model~(\ref{eq:BHH}) with $U/J = 3$, falling into the 
superfluid regime when the system is sufficiently large. Here we restrict 
ourselves to $K_0 = 0$, {\em i.e.\/}, to $\nu = 0$; related studies for 
$\nu = 1,2$ are documented in Ref.~\cite{EckardtHolthaus07}. After selecting 
some frequency $\omega$, and thus specifying the time scale $T = 2\pi/\omega$, 
the time-dependent force is turned on with an amplitude $K_\omega(t)$ 
which rises linearly between $t = 0$ and $t_1 = 100 \, T$ from 
$K_\omega/\hbar\omega = 0$ to $K_{\rm \max}/\hbar\omega = 2.4$. The latter 
value lies close to the first zero of ${\rm J_0}$, and thus gives a quite 
large ratio $U/J_{\rm eff}$, which should place the system far into the 
Mott-like regime. Then the amplitude is kept constant at $K_{\rm \max}$ 
between $t_1$ and $t_2 = 200 \, T$, and finally ramped linearly back to zero 
between $t_2$ and $t_3 = 300 \, T$. The corresponding $N$-body wave function 
$|\psi(t)\rangle$ is computed by plain direct solution  of the time-dependent 
Schr\"odinger equation, not taking any recourse at all to Floquet theory. 
From that wave function, the single-particle quasimomentum distribution  
$\rho(p,t)$ is obtained according to  
\begin{equation}
	\rho(p,t) = \frac{1}{M}\sum_{\ell,j} 
	\exp\!\left[ \ri \frac{(\ell-j)p}{\hbar/a}\right]
	\langle \psi(t) | 
	\hat{b}^\dagger_\ell \hat{b}^{\phantom\dagger}_j
	| \psi(t) \rangle \; , 
\label{eq:QMD}
\end{equation}
and recorded at integer multiples of~$T$~\cite{EckardtEtAl05,EckardtHolthaus07}.
This momentum distribution is sharply peaked, due to (quasi) long-range phase 
coherence, in the superfluid phase, but apparently structureless in the Mott 
regime.

Figure~\ref{F_1} shows the results for $\hbar\omega/J = 16$: Initially one
finds a strongly peaked distribution, as expected for a superfluid-like state, 
which then becomes practically flat at $t = t_1$. However, after the driving 
amplitude $K_\omega(t)$ has been switched off again at $t = t_3$, the sharp
pattern reappears: This fact clearly signals that the flat distribution
between $t_1$ and $t_2$ is not due to loss of coherence resulting from
uncontrolled excitations, but rather indicates the Mott-like regime, since 
otherwise it would not be possible to switch back (almost) adiabatically to 
the initial state. Thus, this figure provides a glimpse at a quantum
phase transition induced by ``dressing'' a matter wave, although, of course,
a truly sharp ``transition'' cannot be achieved with $N = M =7$.

\begin{figure}[t]
\begin{center}
	\includegraphics[angle=-90.0, width = 1\linewidth]{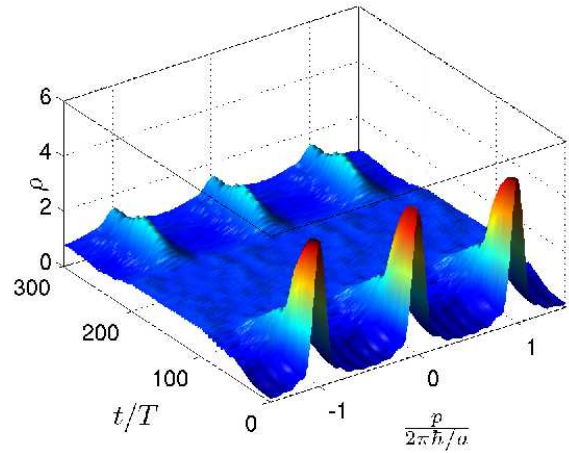}
\end{center}
\caption[FIG.~2]{As Fig.~\ref{F_1}, but for a lower frequency 
	$\hbar\omega/J = 12$. In this case, adiabatic following is destroyed 
	by Landau--Zener transitions at the avoided quasienergy crossings 
	visible in Fig.~\ref{F_4}.}
\label{F_2}
\end{figure}

But the small system already is sufficently rich to demonstrate that 
an ideal outcome cannot be taken for granted: Fig.~\ref{F_2} shows a 
momentum distribution obtained in the same manner for a lower frequency, 
$\hbar\omega/J = 12$. Whereas a signature like this might not be 
distinguishable from that in Fig.~\ref{F_1} in an actual experiment for 
times up to $t_2$, here the initial pattern is not restored at $t_3$, 
indicating severe deviations from the desired adiabatic following. 

\begin{figure}[t]
\begin{center}
	\includegraphics[angle=-90.0, width = 1\linewidth]{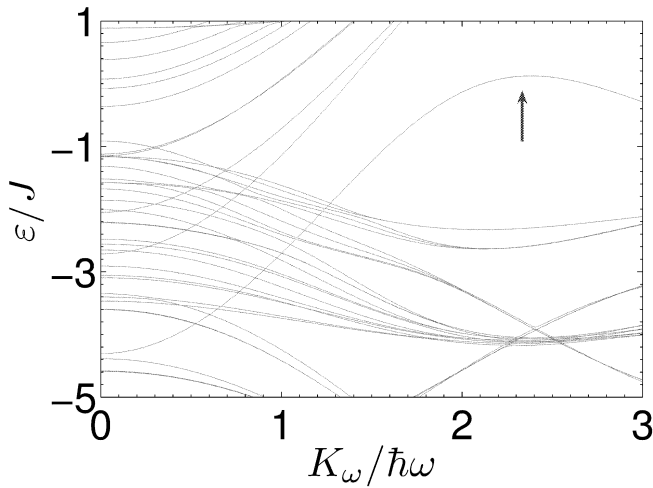}
\end{center}
\caption[FIG.~3]{Part of the quasienergy spectrum for the almost ideal case 
	considered in Fig.~\ref{F_1}. The arrow marks the quasienergy of the 
	Floquet state evolving adiabatically from the ground state of the 
	undriven Bose--Hubbard system. Computed with $N = M =5$.}
\label{F_3}
\end{figure}

\begin{figure}[th]
\begin{center}
	\includegraphics[angle=-90.0, width = 1\linewidth]{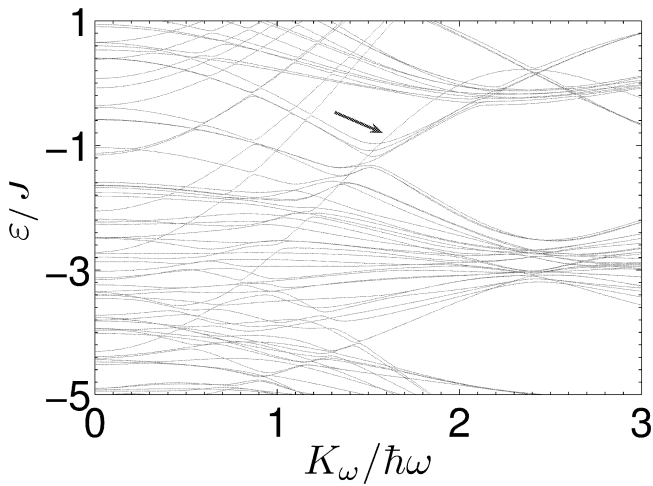}
\end{center}
\caption[FIG.~4]{Part of the quasienergy spectrum ($N = M = 5$) for the 
	thwarted case considered in Fig.~\ref{F_2}; the arrow indicates 
	the quasienergy associated with the ground state. 
	Observe that here 	
	there are several active resonances, corresponding to pronounced 
	avoided level crossings.}	 
\label{F_4}
\end{figure}
 
Inspection of the corresponding quasienergy spectra immediately reveals the 
reason for the different dynamics found in both cases. In Fig.~\ref{F_3} 
we depict a part of the quasienergy spectrum for the frequency employed in 
Fig.~\ref{F_1}, whereas Fig.~\ref{F_4} shows the spectrum for the lower 
frequency underlying Fig.~\ref{F_2}; both spectra have been computed with 
$N = M = 5$ by solving the eigenvalue equation~(\ref{eq:QEE}). In the first 
case, the quasienergy line emanating from the ground state is not visibly 
affected by other states, indicating the absence of active resonances and thus 
enabeling the adiabatic return observed in Fig.~\ref{F_1}; essentially; only 
one single instantaneous Floquet state is populated during the entire process. 
In contrast, in the second case the quasienergy level originating from the 
ground state undergoes several large avoided crossings. Incomplete 
Landau--Zener transitions at these avoided quasienergy crossings then 
lead to a significant population of the anticrossing Floquet 
states~\cite{DreseHolthaus99}, rendering an adiabatic return to the initial 
state impossible.

\section{Conclusions}
\label{S_6}

Ultracold atoms in time-periodically modulated optical lattices give rise 
to ``dressed matter waves'', in analogy to the dressed atoms known from 
atomic physics~\cite{CohenTannoudji04}. Such dressed systems acquire properties
quite different from their ``bare'' antecedents, the modification of atomic 
Land\'e $g$-factors setting a prominent example~\cite{HarocheEtAl70}. An effect
closely related to this $g$-factor modification is a transition of ultracold 
bosonic atoms in a modulated optical lattice from a superfluid to a 
Mott-insulator-like state in response to a variation of the modulation 
strength; this transition is mediated by a modification of the 
nearest-neighbour hopping matrix elements relying on precisely the same 
mechanism as that of the $g$-factors. An experimental verification of this 
proposal involves adiabatic following of the many-body Floquet state 
originating from the ground state of the bare system; such adiabatic following 
is a quite tricky concept in the context of driven matter waves. The question 
to what extent multiphoton-like resonances can be avoided (or perhaps  
deliberately be induced and exploited) is open to experimental investigation.

The long-term perspective of these considerations, however, seems to lie 
elsewhere. Just as the $g$-factor modification is but {\em one\/} facet of 
the dressed-atom picture, there are further possibilites of controlling 
the state of a matter wave in an optical lattice by time-periodic forcing.
Our present scheme defines a first cornerstone; if achieved, more demanding 
ones can follow. In particular, it might be interesting to resonantly couple 
different Wannier states located at the same site, and thus to open up new 
ways of quantum state engineering. The experimentally established fact 
that Bose--Einstein condensates in optical lattices can be subjected to 
strong forcing without destroying their phase coherence~\cite{LignierEtAl07} 
is a sound cause for optimism.

\section*{Acknowledgments}
We thank the participants of the 395th Wilhelm and Else Heraeus Seminar
{\em Time Dependent Phenomena in Quantum Mechanics\/} for stimulating
discussions, and E.~Arimondo, O.~Morsch and their team for introducing us 
to their experiments~\cite{LignierEtAl07,SiasEtAl07}. This work was supported 
in part by the Deutsche Forschungsgemeinschaft through the Priority Programme 
SPP~1116.

%\section*{References}


\begin{thebibliography}{99}

\bibitem{CohenTannoudji04} Cohen-Tannoudji C 2005
	{\it Atoms in Electromagnetic Fields\/}
	(Singapore: World Scientific)
	
\bibitem{HarocheEtAl70} Haroche S, Cohen-Tannoudji C, Audoin C and
	Schermann J P 1970
	{\it Phys. Rev. Lett.} {\bf 24} 861
	
\bibitem{ThomsonEtAl92} Thomson D S, Renn M J and Gallagher T F 1992
	{\it Phys. Rev.} A {\bf 45} 358 

\bibitem{LignierEtAl07} Lignier H, Sias C, Ciampini D, Singh Y, 
	Zenesini A, Morsch O and Arimondo E 2007
	{\it Phys. Rev. Lett.} {\bf 99} 220403

\bibitem{SiasEtAl07} Sias C, Lignier H, Singh Y P, Zenesini A, 
	Ciampini D, Morsch O and Arimondo E 2007 
	arXiv:0709.3137v1 (to appear in {\it Phys. Rev. Lett.})
	
\bibitem{EckardtEtAl05} Eckardt A, Weiss C and Holthaus M 2005
	{\it Phys. Rev. Lett.} {\bf 95} 260404 

\bibitem{EckardtHolthaus07} Eckardt A and Holthaus M 2007
	{\it EPL} {\bf 80} 50004		

\bibitem{CreffieldMonteiro06} Creffield C E and Monteiro T S 2006
	{\it Phys. Rev. Lett.} {\bf 96} 210403
	
\bibitem{AbramowitzStegun72} Abramowitz M and Stegun I A (eds.) 1972 
	{\it Handbook of mathematical functions\/}
	(New York: Dover Publications)
	
\bibitem{GrossmannEtAl91} Grossmann F, Dittrich T, Jung P and H\"anggi P 1991	
	{\it Phys. Rev. Lett.} {\bf 67} 516
	
\bibitem{LlorentePlata92} Gomez Llorente J M and Plata J 1992 
	{\it Phys. Rev.} A {\bf 45} R6958;
	Erratum: 1994 {\it Phys. Rev.} E {\bf 49} 3547
	
\bibitem{Oberthaler07} Oberthaler M K, private communication	
		
\bibitem{AutlerTownes55} Autler S H and Townes C H 1955 
	{\it Phys. Rev.} {\bf 100} 703

\bibitem{Shirley65} Shirley J H 1965
	{\it Phys. Rev.} {\bf 138} B979 

\bibitem{Zeldovich67} Zel'dovich Ya B 1966
    	{\it Zh. Eksp. Theor. Fiz.} {\bf 51} 1492
    	(1967 {\it Sov. Phys. JETP} {\bf 24} 1006)
    
\bibitem{Ritus67} Ritus V I 1966
    	{\it Zh. Eksp. Theor. Fiz.} {\bf 51} 1544 
    	(1967 {\it Sov. Phys. JETP} {\bf 24} 1041)

\bibitem{Sambe73} Sambe H 1973
	{\it Phys. Rev.} A {\bf 7} 2203 
	
\bibitem{BornFock28} Born  M and Fock V 1928 
    	{\it Z. Phys.} {\bf 51} 165

\bibitem{Berry84} Berry M V 1984
    	{\it Proc. R. Soc. Lond.} A {\bf 392} 45 

\bibitem{BreuerHolthaus89a} Breuer H P and Holthaus M 1989
	{\it Z. Phys.} D {\bf 11} 1
	
\bibitem{BreuerHolthaus89b} Breuer H P and Holthaus M 1989
	{\it Phys. Lett.} A {\bf 140} 507 
		
\bibitem{DreseHolthaus99} Drese K and Holthaus M 1999
	{\it Eur. Phys. J.} D {\bf 5} 119 		
	
\bibitem{HoneEtAl97} Hone D W, Ketzmerick R and Kohn W 1997
	{\it Phys. Rev.} A {\bf 56} 4045
	
\bibitem{FisherEtAl89} Fisher M P A, Weichman P B, Grinstein G and 
	Fisher D S 1989 
	{\it Phys. Rev.} B {\bf 40} 546

\bibitem{KuehnerEtAl00} K\"uhner T D, White S R and Monien H 2000
	{\it Phys. Rev.} B {\bf 61} 12474 

\bibitem{JakschEtAl98} Jaksch D, Bruder C, Cirac J I, Gardiner C W 
	and Zoller P 1998
	{\it Phys. Rev. Lett.} {\bf 81} 3108

\bibitem{GreinerEtAl02} Greiner M, Mandel O, Esslinger T, H\"ansch T W 
	and Bloch I 2002
	{\it Nature} {\bf 415} 39

\bibitem{StoeferleEtAl04} St\"oferle T, Moritz H, Schori C, K\"ohl M 
	and Esslinger T 2004
	{\it Phys. Rev. Lett.} {\bf 92} 130403

\bibitem{MunEtAl07} Mun J, Medley P, Campbell G K, Marcassa L G, 
	Pritchard D E and Ketterle W 2007
	{\it Phys. Rev. Lett.} {\bf 99} 150604
	
\bibitem{BlochEtAl07} Bloch I, Dalibard J, and Zwerger W 2007
	arXiv:0704.3011v2 (to appear in {\it Rev. Mod. Phys.})
		
\end{thebibliography}
\end{document}